# Observed transition from Richtmyer-Meshkov jet formation through feedout oscillations to Rayleigh-Taylor instability in a laser target

Y. Aglitskiy,[1] M. Karasik,[2] A. L. Velikovich,[2] V. Serlin,[2] J. Weaver,[2] T. J. Kessler,[2] S. P. Nikitin,[3] A. J. Schmitt,[2] S. P. Obenschain,[2] N. Metzler,[3,4] and J. Oh[3]

[1]*Science Applications International Corporation, McLean, VA 22150*
[2]*Plasma Physics Division, Naval Research Laboratory, Washington, DC 20375*
[3]*Research Support Instruments, Lanham, MD 20706*
[4]*Ben Gurion University, Beer Sheva, Israel*

Experimental study of hydrodynamic perturbation evolution triggered by a laser-driven shock wave breakout at the free rippled rear surface of a plastic target is reported. At sub-megabar shock pressure, planar jets manifesting the development of the Richtmyer-Meshkov-type instability in a non-accelerated target are observed. As the shock pressure exceeds 1 Mbar, an oscillatory rippled expansion wave is observed, followed by the "feedout" of the rear-surface perturbations to the ablation front and the development of the Rayleigh-Taylor instability, which breaks up the accelerated target.

PACS numbers: 52.57.Fg, 52.70.La, 52.35.Tc, 47.20.Ma

When a planar shock wave hits a rippled interface between two different materials, it triggers the development of the classical Richtmyer-Meshkov (RM) instability [1]. This instability develops for any density ratio across the interface and for arbitrary equations of state (EOS) of the shocked materials, provided that the shock pressure is sufficiently high to overcome the material strength. It is driven mainly by the surface vorticity deposited at the interface by shock refraction [2].

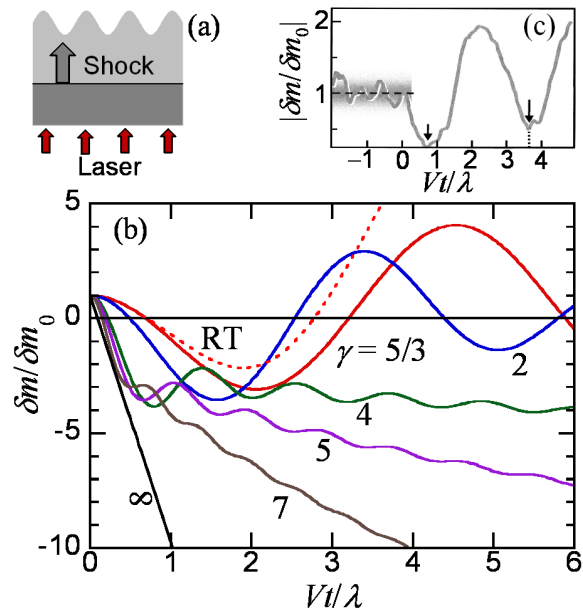

FIG. 1 (a) Schematic of the experiment: laser-driven planar shock wave approaches rippled free rear surface of a target. (b) Solid lines: Normalized areal mass modulation amplitude vs. time for a rippled expansion wave in a very thick target. Dotted line: Same for a finite-thickness target, where the RT growth is triggered by the feedout mechanism. (c) Experimental data from [9] roughly corresponding to the dotted RT line in (b).

What happens when a planar shock wave breaks out at a free rippled surface as shown in Fig. 1(a)? This is not a clear case of the RM instability, since no shock refraction occurs and no surface vorticity is deposited anywhere. The classical RM instability develops independently of the EOS, but this is not the case here. If the shock is strong enough for complete vaporization of the shocked material on unloading [3], then, as shown in Ref. [4], the shock breakout is followed by the formation of an oscillating rippled expansion wave that propagates back into the shocked gas without any instability growth. On the other hand, it is well known that when the material remains incompressible in the shocked state, and if its rear surface is initially indented, then the shocked indentation inverts itself, forming a jet (the shaped charge effect), which closely resembles the development of the RM instability. This has been theoretically predicted and consistently observed in both explosively- and laser-driven experiments, as well as in the simulations, see Refs. [5-7] and references therein.

In both cases we deal with the same physical phenomenon, namely, the evolution of a rippled expansion wave after shock breakout at a rippled rear surface. But in contrast with the evolution of the shocked rippled material interface, i. e. the classical RM case, a rippled expansion wave behavior can change qualitatively when the EOS is varied. This property, first noticed in Ref. [4], has been studied in detail in Ref. [8], the relevant results of which are summarized in Fig. 1(b). The time histories of the areal mass modulation amplitude, $\delta m$, are shown for several finite values of ideal-gas $\gamma$, and in the limit $\gamma \to \infty$ modeling an incompressible fluid. Here, $\delta m$ is normalized with respect to its initial value $\delta m_0 = \rho_0 \delta x_0$, where $\rho_0$ and $\delta x_0$ are, respectively, the initial values of the material density and the ripple amplitude. The time is normalized with respect to $\lambda/V$, where $\lambda$ is the ripple wavelength (the standard small-amplitude assumption $\delta x_0 \ll \lambda$ is made), $V = 2a_s/(\gamma-1)$ is the velocity with which the expansion front propagates into vacuum, $a_s$ is the post-shock speed of sound. The origin $t=0$ in Fig. 1(b) is the instant of shock breakout at the rippled surface; the distance separating it from the ablation front at $t=0$ is supposed to be much greater than $\lambda$. The solid lines plotted for finite $\gamma$ are exact small-amplitude solutions of [8]. The late-time evolution of $\delta m$ depends on whether the gas $\gamma$ is less or greater than 3. In the former case, the evolution is oscillatory; in the latter, a secular power-law growth dominates:

$$\frac{\delta m}{\delta m_0} \cong \begin{cases} \dfrac{4\sqrt{2}\left[\gamma-1+\sqrt{2\gamma(\gamma-1)}\right]\sin\Omega\tau}{(\gamma-1)^2 \Omega\tau}, & \gamma<3; \\ -\dfrac{2\left[2(\gamma-1)^2+(\gamma-3)\sqrt{2\gamma(\gamma-1)}\right]}{(\gamma-1)^4}\Gamma\!\left(\dfrac{2}{\gamma-1}\right)\!\left(\tau\sqrt{\dfrac{\gamma+1}{\gamma-3}}\right)^{(\gamma-3)/(\gamma-1)}, & \gamma>3, \end{cases} \quad (1)$$

where $\Omega = [2(\gamma-1)/(\gamma+1)]^{(\gamma-1)/(3-\gamma)}$ and the time variable $\tau = \pi(\gamma-1)Vt/\lambda$ is assumed to be much greater than unity. In the incompressible limit $\gamma \to \infty$, the value of $\tau$ tends to infinity for any positive value of normalized time $Vt/\lambda$. Constant density $\rho = \rho_0$ allows one to express the time-dependent ripple amplitude $\delta x(t)$ via $\delta m$ in this limit from (1):

$$\delta x(t) = \lim_{\gamma \to \infty} \frac{\delta m}{\rho_0} = \delta x_0 - \left(1+\frac{1}{\sqrt{2}}\right)k\delta x_0 Vt, \qquad (2)$$

where $k = 2\pi/\lambda$, and the initial condition $\delta x(0) = \delta x_0$ has been taken into account. This is very similar to the linear growth predicted by the Richtmyer's formula [1], which is exact for an incompressible fluid impulsively accelerated to the velocity $V$ in the small-amplitude approximation. But the growth rate of $\delta x(t)$ in (2) is seen to be higher, by a factor of $1+1/\sqrt{2} = 1.707$, than $-k\delta x_0 V$, which is predicted by the Richtmyer's formula at Atwood number $A = -1$ appropriate for a free rear surface.

The first change of sign of $\delta m$ in Fig. 1(b) means a reversal of phase in the areal mass distribution: more mass gets accumulated in the valleys, where there was less of it initially. For low $\gamma$, the phase of $\delta m$ in a sufficiently thick target reverses itself many times, cf. Eq. (1). For high $\gamma$ the phase reversal occurs just once, after which the areal mass keeps accumulating in the valleys. In a laser-driven target of finite thickness, the situation changes when the leading edge of the rippled expansion wave breaks out at the ablation front directly opposite from the locations of the valleys. At this instant, the target starts accelerating, and the Rayleigh-Taylor (RT) instability growth is triggered [9, 10]. The growing bubbles lose their mass to the spikes, located where the target was initially thicker. The direction of variation of $\delta m$ thus becomes positive, $\delta m$ reverses its phase again and keeps growing, as illustrated by the dotted line labeled RT plotted for $\gamma = 5/3$ in Fig. 1(b). Our earlier direct experimental observation of such behavior with face-on radiography [9] is illustrated by Fig. 1(c), where the thickness of the darkened area approximately corresponds to the experimental uncertainty and the vertical arrows mark the instants of phase reversals.

Dynamics of the expansion wave is governed by the release isentrope of the shocked material, a single line in the thermodynamic phase space of the shocked material [3]. The adiabatic index defined for the isentropic expansion as $\gamma = \left(\partial \ln p / \partial \ln \rho\right)_S$ changes along the isentrope for realistic EOS. The approximation assuming $\gamma$ to be constant along the release isentrope is very rough. Still, it is valid in the two limiting cases: a) a relatively weak shock, leaving behind an almost incompressible shocked material, in which large pressure variation corresponds to a small change of density, $\gamma \to \infty$; b) a very strong shock that turns the shocked material into a gas thermodynamic state, $\gamma \to 5/3$. The goal of this work is to observe the qualitative transition from the incompressible (RM-like growth) to the fully compressible (oscillations) perturbation evolution. To do this, one needs to gradually increase the shock pressure from sub-Mbar to multi-Mbar in the same material, in the same experimental setting.

This is difficult to do with explosive drive [6] but possible when the shock waves are driven by the laser ablation pressure. Our experiments were done on the 56 beam krypton fluoride (KrF) Nike laser facility with laser wavelength $\lambda_L = 248$ nm and a 4 ns pulse. We varied the number of beams overlapped on the plastic target to change the ablative pressure: 36 beams produce ablative pressure of ~8 Mbar, whereas a single beam irradiation reduces the pressure to ~0.7 Mbar.

A monochromatic x-ray imaging system based on Bragg reflection from a spherically curved quartz crystal [9, 11, 12] was used to radiograph the main target. The crystal selects the He-like Si (1.86 keV) resonance line of a Si backlighter and projects a monochromatic image of the target on an x-ray detector.

To observe thin planar jets side-on we had to ensure that both the target plane and the ripples were parallel to the diagnostic line of sight with the accuracy of 5 mrad and to maintain the Bragg angle at the same time. To avoid including peripheral areas of the laser spot into the side-on image, the targets were narrow strips: about 500 μm wide, only slightly larger than the 400 μm diameter flat-top of Nike's smooth focal spot. We used 53 μm thick polystyrene targets with a laser machined single-mode ripple wavelengths of $\lambda = 46$ μm and initial peak-to-valley amplitude of $2\delta x_0 = 15$ μm on the rear side.

To ensure a near-perfect correlation between the initial rear surface ripples and the evolved perturbations we aligned and overlapped the images from a backlighter-only pre-shot and the driven images of the same target. To resolve the positions of the evolved structures with respect to the initial rear-surface ripples we needed the images of a substantial number of ripples from the undisturbed parts of the target to be well within our field of view, which thereby had to be significantly larger than the laser spot size (750 µm FWHM) while maintaining ~10µm spatial resolution. With high magnification of 15× it made an imaging plate a practical choice for the detector.

The temporal resolution was achieved by using short, 400 ps to 1 ns, x-ray pulses from the backlighter beams that could be synchronized with the desired stage of the process under investigation. When the drive energy was low, about 40 J, the backlighter time was varied between 2 and 17 ns after the start of the drive pulse. When the drive energy was above one kJ, target evolution was occurring much faster and the images were taken within several ns of the drive pulse start. For low energy shots, or at the early stages of the high energy shots, while the target motion was not too fast, our backlighter pulses were short enough to make all the details of the perturbation evolution clearly visible. At the acceleration stage of the high-energy shots, substantial motion blurring is observed in the images.

We have also used a stroboscopic imaging mode, in which the backlighter beams had two consecutive pulses of 400 ps separated by a 2.4 ns interval. The two x-ray flashes from the same backlighter target produced two overlapped, and yet clearly distinct, images of the evolving main target on the same imaging plate.

Figure 2(a) shows an image taken 15 ns after the start of a low energy (single-beam) drive pulse, laser intensity on the target being ~$1.4 \times 10^{12}$ W/cm$^2$. At the time of observation, 11 ns have passed since the end of the laser pulse, and at least 8 ns since the decompression of the target's rear surface. Still, a very regular array of narrow, strongly collimated planar jets is observed. The irradiated surface of the target remains planar because the laser pulse is over before the arrival of the head of the rippled expansion wave to the ablation front.

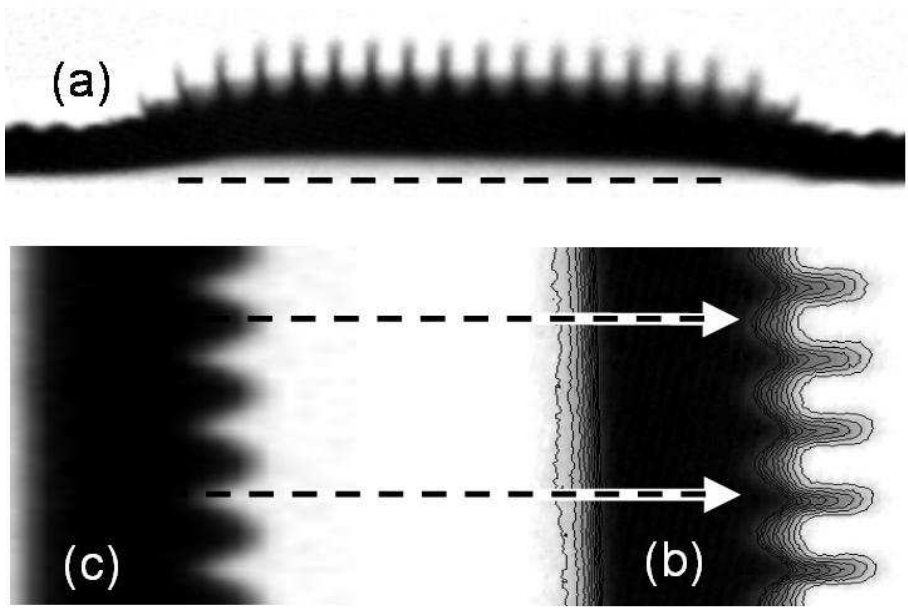

FIG. 2 (a) Radiographic image of a target at 15 ns driven by a single beam. Laser radiation is incident normally on the flat front side of the target. Dashed line marks the initial position of this surface. (b) Density contours of a central part of this target extracted from the image of (a). (c) Backlighter-only image of the same part of this target prior to the shot.

The observed jet formation represents a development of an RM-type instability. For our experimental conditions, the initial value of the nonlinearity parameter is $k\delta x_0 = 1$, implying that the perturbation growth starts from the strongly nonlinear phase. Still, the remarkable regularity of our images proves that the jets are two-dimensional (2D) planar sheets, parallel to each other and perpendicular to the image plane. No random motion indicating transition to turbulence is seen here. The near-perfect uniformity of the observed jet array in Fig. 2(a) in the horizontal direction within the Ø400 μm flat-top of the focal spot attests to the high uniformity of a single beam on Nike.

In Fig. 2(b) we present the density contours extracted from the x-ray image of a few jets in the central part of the target, showing density levels from 0.01 to 0.1 g/cm$^3$ with a 0.01 g/cm$^3$ increment. The estimates have been made using a cold polystyrene opacity 345 cm$^2$/g of the 1.86 keV x rays and assuming an equivalent thickness of the target along the line of sight of 500 μm, the target width. The density contours aligned with the image of the same part of the same target before the shot, Fig. 2(c), demonstrate that these jets flow from the valleys of the rear-surface ripple pattern. We expect this data to be helpful for validation, in the 2D regime, of the hydrodynamic codes in the warm-dense-matter pressure range below 1 Mbar. It requires accurate modeling of the EOS, material strength [6] and other physics once regarded beyond the scope of ICF. There are, however, indications that jets of cold capsule material formed at the early stages of its irradiation propagate from the inner layers of the capsule into the fuel, thereby adversely affecting the ignition [13]. This finding stimulated recent indirectly-driven experiments with bumped plastic targets at about the same shock pressure of ~0.7 Mbar [14] as in our shot illustrated by Fig. 2.

Having slightly increased the laser intensity, we start accelerating the target, and the RT instability comes into play. Figure 3 shows the image taken 16.5 ns after the start of an intermediate energy drive pulse composed of 3 overlapping Nike beams. The peak laser intensity at the center of the target, where all three beams overlap, is ~$4\times10^{12}$ W/cm$^2$, gradually decreasing to the periphery. At this peak intensity there is just enough time in the laser pulse for the head of the expansion wave to reach the ablation front, where the grooves begin to emerge, indicating the onset of the RT instability growth. They are aligned with the rear-side valleys, as they should be. This process of the rear-surface perturbations feeding out to the ablation front has been studied theoretically [10, 4], but this is the first time when the emergence of these grooves at the ablation front is directly observed. Shortly after that the laser pulse is over, so there is not enough time for the RT bubbles to cut through the target. On the rear side we see a manifestation of the standing wave oscillations characteristic of a rippled expansion wave. As the periphery of the target is approached, this pattern gradually morphs into a jet array resembling that of Fig. 2.

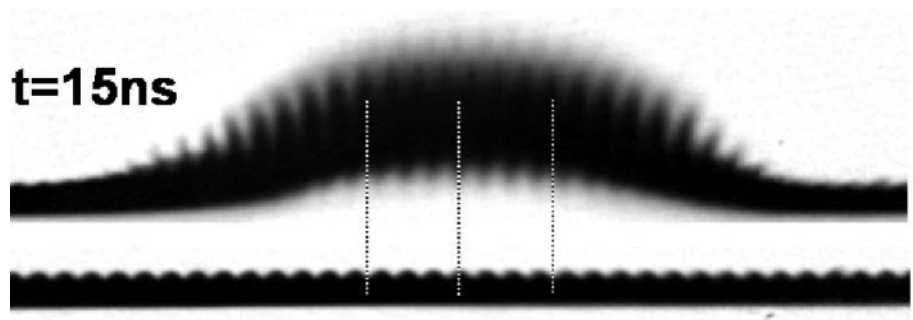

FIG. 3 Radiographic images of a target driven by 3 Nike beams: prior to the main pulse (backlighter only, bottom), and after the end of the laser pulse, at $t=15$ ns (top).

Figure 4 shows the images obtained in two high energy shots, with peak intensity of $5\times10^{13}$ W/cm$^2$ at the center of the target. The left part of Fig. 4 shows two overlapping (stroboscopic) images taken, respectively, at the beginning ($t=1.4$ ns) and near the end ($t=3.8$ ns) of the target acceleration. The earlier image shows the grooves at the ablation front aligned with the valleys on the rear side, same as in Fig. 3. The peaks seen at the rear side of the target on this image are also aligned with the original valleys. They represent the compressible counterparts of the jets observed in Fig. 2, manifesting the oscillation that produces the early-time phase reversal of the areal mass distribution shown in Fig. 1. The later image shows the stage where the RT growth dominates, cf. the dotted RT line in Fig. 1. Here the peaks and valleys both at the rear and front parts of the target are aligned, respectively, with the initial rear-surface peaks and valleys. The phase of the areal mass modulation is positive again. The initial peak areas give rise to the spikes in which the target mass is left; the valley areas – to the bubbles that cut through the target where it was initially thinner. The later stage of such target evolution is shown in the right part of Fig. 4, taken at $t = 15$ ns. The target, having been cut by the RT instability while the laser pulse was on, coasts and freely expands. The expansion reduces the speed of sound, freezing the perturbation structure in it. Even at this late stage of the instability development the image exhibits the imprint of the original ripple pattern and shows no sign of randomness. At its periphery, where the intensity was low and the target did not accelerate, we observe jet formation similar to that of Figs. 2 and 3.

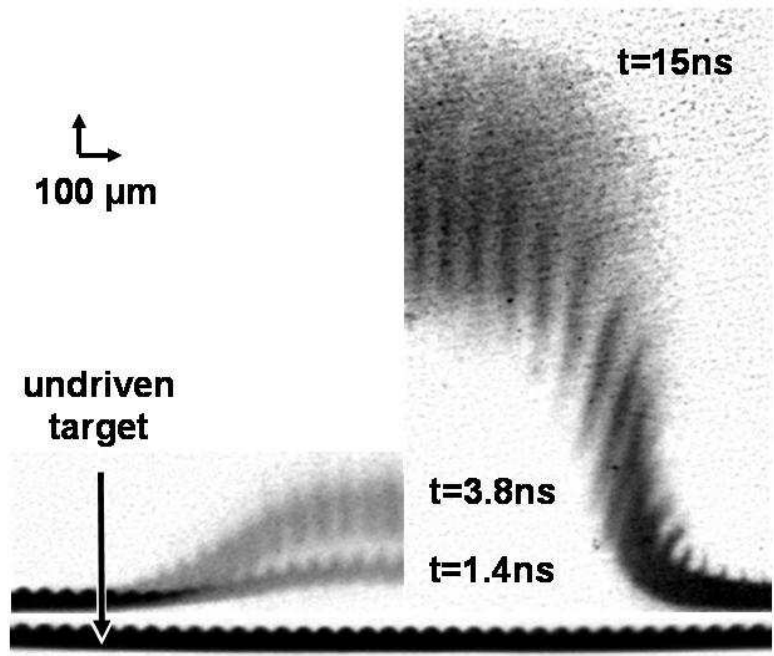

FIG. 4 Left: Radiographic images of a target driven by 36 Nike beams prior to the main pulse (backlighter only), and at the beginning and near the end of its acceleration. Right: Image of a similarly driven target taken with a delay of 15 ns.

To summarize, all the types of evolution of a free rippled rear surface triggered by a strong shock wave, from a shaped-charge jet formation manifesting an RM-type instability to the standing-wave sonic oscillations, are governed by a similar dynamics of a rippled expansion wave and differ only by the EOS of the shocked material [8]. We report the first direct observation of all these types of evolution in the same experimental setting, with the EOS gradually varied from almost incompressible to fully compressible by increasing the intensity of laser radiation that generates the shock wave.

The authors acknowledge the excellent technical support of the Nike Laser Crew. This work was supported by the U. S. Department of Energy, Defense Programs.